\documentclass[acmsmall]{acmart}

\def\BibTeX{{\rm B\kern-.05em{\sc i\kern-.025em b}\kern-.08emT\kern-.1667em\lower.7ex\hbox{E}\kern-.125emX}}
    
\copyrightyear{2018}
\acmYear{2018}
\setcopyright{acmlicensed}
\acmConference[Woodstock '18]{Woodstock '18: ACM Symposium on Neural Gaze Detection}{June 03--05, 2018}{Woodstock, NY}
\acmBooktitle{Woodstock '18: ACM Symposium on Neural Gaze Detection, June 03--05, 2018, Woodstock, NY}
\acmPrice{15.00}
\acmDOI{10.1145/1122445.1122456}
\acmISBN{978-1-4503-9999-9/18/06}

\usepackage{wrapfig}

\begin{document}

%
\title{Human-AI Collaboration in Data Science:  Exploring Data Scientists' Perceptions of Automated AI}

%
\author{Dakuo Wang} 
\authornote{Equal contributions from Dakuo Wang and Justin D. Weisz.}

\email{Dakuo.Wang@ibm.com}
\affiliation{%
  \institution{IBM Research}
}

\author{Justin D. Weisz}
\email{jweisz@us.ibm.com}
\affiliation{%
  \institution{IBM Research}
}

\author{Michael Muller}
\email{michael_muller@us.ibm.com}
\affiliation{%
  \institution{IBM Research}
}

\author{Parikshit Ram}
\email{Parikshit.Ram@ibm.com}
\affiliation{%
  \institution{IBM Research}
}

\author{Werner Geyer}
\email{Werner.Geyer@us.ibm.com}
\affiliation{%
  \institution{IBM Research}
}

\author{Casey Dugan}
\email{cadugan@us.ibm.com}
\affiliation{%
  \institution{IBM Research}
}

\author{Yla Tausczik}
\email{ylatau@umd.edu}
\affiliation{%
  \institution{University of Maryland}
}
\author{Horst Samulowitz}
\email{samulowitz@us.ibm.com}
\affiliation{%
  \institution{IBM Research}
}
\author{Alexander Gray}
\email{Alexander.Gray@ibm.com}
\affiliation{%
  \institution{IBM Research}
}

%
\renewcommand{\shortauthors}{Dakuo Wang et al.}

%
\begin{abstract}
The rapid advancement of artificial intelligence (AI) is changing our lives in many ways. One application domain is data science. New techniques in automating the creation of AI, known as AutoAI or AutoML, aim to automate the work practices of data scientists. AutoAI systems are capable of autonomously ingesting and pre-processing data, engineering new features, and creating and scoring models based on a target objectives (e.g. accuracy or run-time efficiency). Though not yet widely adopted, we are interested in understanding how AutoAI will impact the practice of data science. We conducted interviews with 20 data scientists who work at a large, multinational technology company and practice data science in various business settings. Our goal is to understand their current work practices and how these practices might change with AutoAI. Reactions were mixed: while informants expressed concerns about the trend of automating their jobs, they also strongly felt it was inevitable. Despite these concerns, they remained optimistic about their future job security due to a view that the future of data science work will be a collaboration between humans and AI systems, in which both automation and human expertise are indispensable.
\end{abstract}

%
%

\ccsdesc[500]{Human-centered computing~Empirical studies in HCI}
\ccsdesc[300]{Computing methodologies~Artificial intelligence}
\ccsdesc[300]{Computing methodologies~Machine learning}

%
\keywords{AutoAI, Human-AI Collaboration, data science, automation, automl, data scientist, domain experts, future of work}

%

\setcopyright{acmlicensed}
\acmJournal{PACMHCI}
\acmYear{2019} \acmVolume{3} \acmNumber{CSCW} \acmArticle{211} \acmMonth{11} \acmPrice{15.00}\acmDOI{10.1145/3359313}

%
\maketitle

\section{Introduction}
Data science is a multi-disciplinary field focused on generalizable processes for extracting knowledge and insights from data~\cite{Dhar:2013}. Data scientists often integrate a wide range of skills from domains including mathematics and statistics, machine learning and artificial intelligence, databases and cloud computing, and data visualization. In recent years, data science has become increasingly popular as organizations embrace data-driven decision-making approaches. However, organizations are struggling to recruit enough data scientists~\cite{Markow:2017}. Thus, employees with a background in software engineering or business analytics typically transfer to data scientist roles when data science teams are formed. In addition, some organizations (e.g. non-profit organizations) rely on citizen data scientists to help analyze non-proprietary data in events such as hackathons~\cite{hou2017hacking} or online challenges~\cite{web:kaggle}. These models, while often successful, are not without their own challenges~\cite{muller2019datascience}.

In addition to operational innovations in conducting data science, researchers and practitioners have developed new kinds of tools to improve the efficiency of data science work. Jupyter Notebook~\cite{web:jupyter}, its successor JupyterLab~\cite{web:jupyterlab}, and Google Colab~\cite{web:colab}, are notable coding environments being used widely in the data science community. Each environment provides a space in which developers can write both code and narratives (e.g., descriptions of data, code, or visualizations) in cells next to each other. Data scientists find it particularly useful in their efforts to ``craft the data'' and tell a story from the data~\cite{muller2019datascience}. HCI researchers have spent much effort in understanding the work practices of data scientists and how to build new tools and features to support them (e.g.~\cite{muller2019datascience, hou2017hacking, passi2017data, heer2019agency, heer2007voyagers, gil2019towards}). For example, Heer et al. developed Data Wrangler~\cite{kandel2011wrangler} and Voyager~\cite{heer2007voyagers} to support data scientists with data pre-processing tasks and exploratory visual analysis tasks, respectively.

In parallel to the development of new features and tools to increase work efficiency, another thread of technical development focuses on the creation of technologies that automate tasks within the data science workflow. These systems, which we collectively refer to as ``AutoAI,'' are being developed by numerous players, including large technology companies, startups, and open source projects. Examples include Google's AutoML~\cite{web:googleautoml}, 
H2O~\cite{web:h2o}, DataRobot~\cite{datarobot}, and open source libraries such as Auto-sklearn~\cite{feurer2015efficient} and TPOT~\cite{olson2016tpot,web:tpot}. Many of these systems build upon the widely used \texttt{scikit-learn} Python machine learning library~\cite{scikit_learn}. AutoAI systems, if their underlying technologies are truly able to automate the data science workflow, may be instrumental in bridging the gap between the high demand and low supply of data scientists, as well as enabling existing data science teams to operate more productively.

But, how do data scientists feel about a future in which their jobs have largely been automated? We seek to understand how data scientists perceive AutoAI technology and how it might be incorporated into or displace aspects of their work practice. To address this question, we conducted interviews with 20 data scientists at IBM, all of whom practice data science at a high technical level in various business contexts. We acknowledge that the term ``data scientist'' often refers to a broad class of people who conduct tasks both directly and indirectly related to data science (e.g. information strategy, data governance, ethics). Muller et. al proposed the term ``data science worker'' to capture this broad diversity~\cite{muller2019datascience}. In this paper, we focus only on self-described ``data scientists'' -- people who actively practice the craft of data science -- and use that term accordingly.

Our results suggest that data scientists have mixed feelings toward AutoAI. Some dislike the high level of automation and feel it will be detrimental to their craft. Others welcome AutoAI as a collaborator and teacher who will ease the burden of labor-intensive tasks and help implement better practices. Yet, nearly all interviewees felt that ``AutoAI is the future of data science,'' and that it would have a significant impact on the field.

Our paper makes the following contributions to the CSCW literature:

\begin{enumerate}
    \item We provide one of the first examinations of AutoAI technologies through the lens of data science practitioners. We identified a ``scatter-gather'' pattern of collaboration amongst data scientists, in which collaborative work and individual work take place in alternating cycles. During \textit{scatter} phases, AutoAI can improve the productivity of an individual data scientist by automating aspects of data cleaning, feature engineering, and model building. By contrast, during \textit{gather} phases, when collaboration happens at the level of ideas rather than artifacts, AutoAI can advise teams about analyses to perform based on the efforts of other team members.
    \item We characterize AutoAI as not just another ``tool'' in the toolbox of individual data science practitioners, but as a first-class subject in data science practice capable of collaborating across a team of data scientists and other stakeholders, thus existing CSCW understandings may be applied to design this AI system's collaboration with human. While the focus of AutoAI development is currently on applying AI techniques to automate the creation of AI models, AutoAI can also be shaped to help build consensus amongst stakeholders via transparency, recommendation, and explanation.
    \item Data scientists themselves see AutoAI as taking on a complementary role to their own, never eliminating the need for their own human input in data science work. Nonetheless, other stakeholders (e.g. data science managers) are attracted by the potential cost savings associated with automating portions of the data science pipeline.
\end{enumerate}

We begin with a short introduction about the technical foundations of AutoAI and show an example of an AutoAI user interface. We next review relevant HCI and CSCW literature in data science. After reporting our methods and results, we discuss the different roles that AutoAI can take in the data science process. We conclude by envisioning a collaborative future between humans and AI, enabled by AutoAI.

\begin{figure}[ht]
  \begin{center}
    \includegraphics[width=0.8\textwidth]{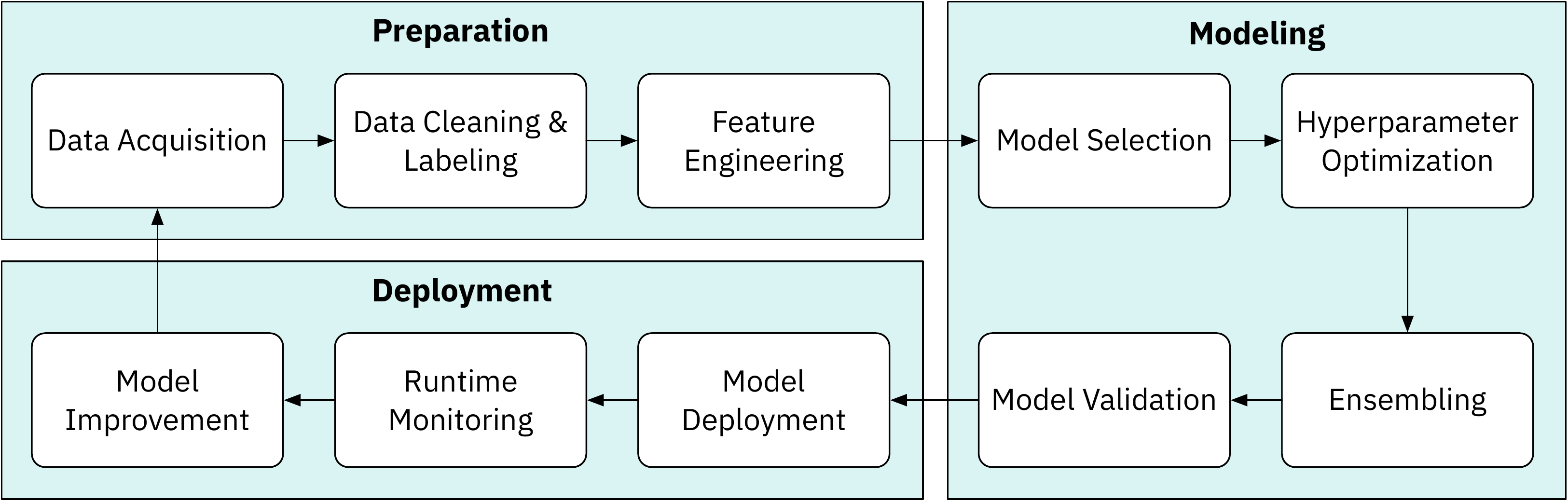}
  \end{center}
  \caption{Typical steps of a data science workflow, consisting of three high-level phases: data preparation, model building, and model deployment.}
  \label{fig:DS-steps}
\end{figure}

\section{What is AutoAI?}
Data science is an iterative, labor-, and time-intensive process. Figure~\ref{fig:DS-steps} shows one view of the data science workflow, common to AutoAI pipelines. It begins with the acquisition, cleaning, and labeling of data, then moves to engineering features, building models, deploying and monitoring models, and collecting feedback that can be used to improve the system. \emph{Data acquisition} is the process of acquiring data that will be used to build a model. \emph{Data cleaning \& labeling} is the process by which data are cleaned (e.g. deciding how to handle missing values or how to encode categorical features) and labeled (e.g. using human annotators to provide ground truth). \emph{Feature engineering} is the process of splitting, combining, and/or transforming columns of data to produce more meaningful features. \emph{Model selection} is the process of choosing the right model to fit to the data (e.g. choosing a random forest to perform a classification task). \emph{Hyperparameter optimization} is the process of choosing model parameters that make it perform well (e.g. how many trees to create in a random forest). \emph{Ensembling} refers to the practice of training multiple models and combining their results to build a more powerful model. \emph{Model validation} is the process of ensuring the generalizability of a model. \emph{Model deployment} happens when a model is put into practice, and \emph{runtime monitoring} ensures the model performs adequately on real data. Feedback collected from a deployed model is used for \emph{model improvement}.

Much work in the fields of AI and machine learning has been conducted over the past few years to automate each of these steps. For \emph{data cleaning}, Kougka et al. provide a survey of different techniques that can be used to automate the data cleaning process (also known as ETL or Extract/Transform/Load)~\cite{kougka2018many}. \emph{Feature engineering} is often considered to be the most time consuming step of the data science process~\cite{Kaggle2017}, and techniques such as Deep Feature Synthesis~\cite{kanter2015deep} and One Button Machine~\cite{lam2017one} automate the task of generating ML-ready features from multi-relational databases. There are also techniques that try to \emph{engineer new features} for a target ML model using reinforcement learning~\cite{khurana2016cognito} and guided search based on past experiences~\cite{nargesian2017learning}. Tools such as AutoWEKA~\cite{autoweka1,autoweka2}, Auto-sklearn~\cite{feurer2015efficient}, and TPOT~\cite{web:tpot,olson2016tpot} all automate the \emph{model selection}, \emph{hyperparameter optimization}, and \emph{ensembling} steps of the data science pipeline. With the rising popularity of deep learning methods, finding the right combination of model architecture and hyperparameters becomes even more critical for building well-performing models. Techniques such as Bayesian optimization~\cite{shahriari2016taking} have been developed to automate hyperparameter optimization, and reinforcement learning and evolutionary algorithms have both been applied to the problem of automating the design of neural architectures~\cite{elsken2018neural}. Taken together, these techniques are helping shape the field of \emph{meta-learning}, the science of learning about how different machine learning approaches perform on a wide range of learning tasks~\cite{vanschoren2018meta}.  But, these efforts are scattered, and users need an end-to-end solution to leverage these techniques. Thus, start-up companies such as H2O~\cite{web:h2o} and DataRobot~\cite{datarobot} are building enterprise data science tools that also automate the \emph{model deployment} and \emph{runtime monitoring} phases of the data science pipeline. 

\subsection{AutoAI Demo}
\label{sec:autoai_demo}
\begin{figure}[ht]
  \begin{center}
    \includegraphics[width=0.8\textwidth]{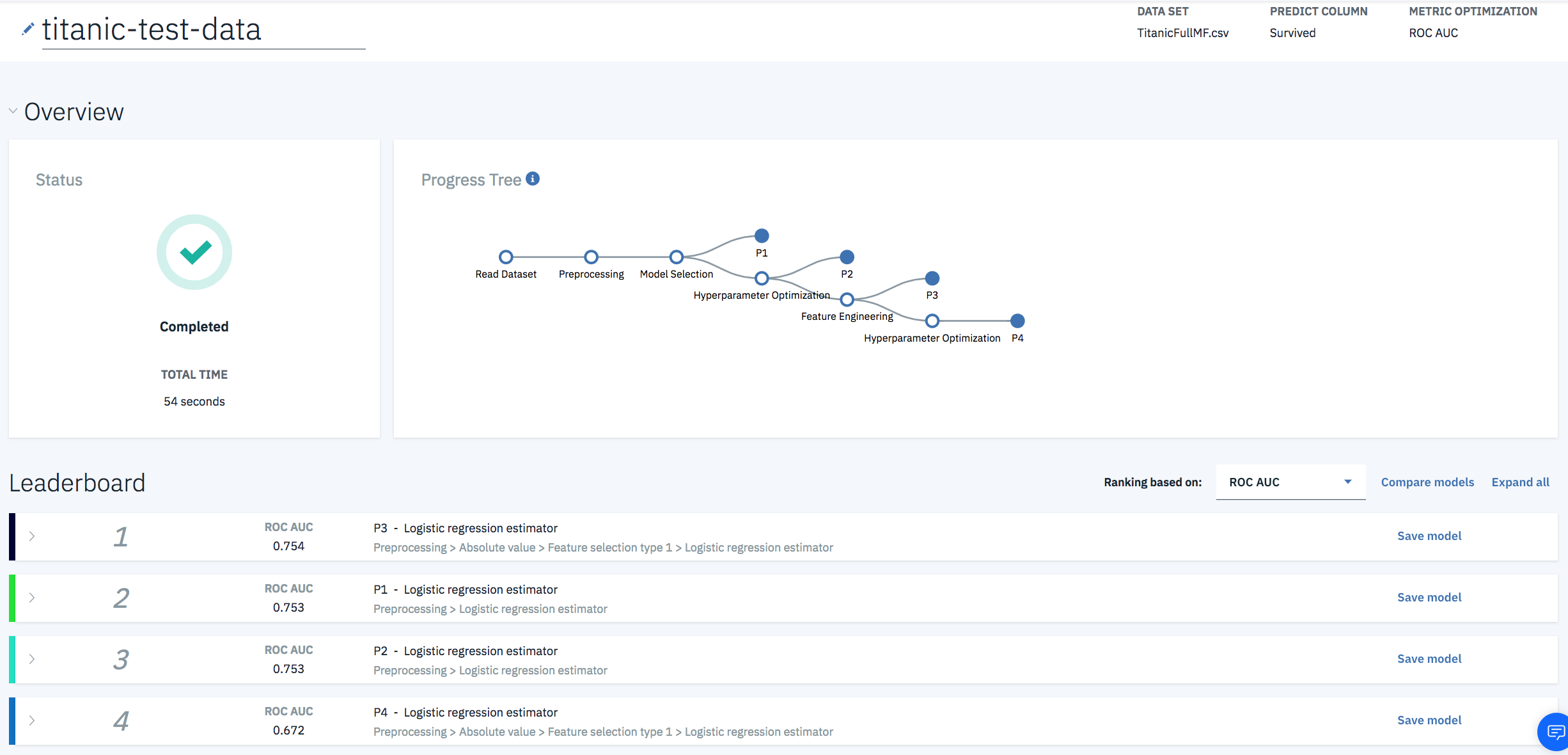}
  \end{center}
  \caption{Screenshot from the AutoAI demo used in our interview study. Top-left: progress pane showing the status of the underlying AutoAI compute job. Top-right: Visualization of the operations performed to create a set of candidate models. Bottom: Leaderboard ranks models based on a desired metric (e.g. ROC AUC, F-1 score, accuracy, etc.)}
  \label{fig:autoai-ui}
\end{figure}

We have built a demo that showcases one possible AutoAI workflow for data scientists. We show an example screenshot from this demo in Figure~\ref{fig:autoai-ui}, which shows the state of the UI after the AutoAI process has completed. The UI consists of three main sections: 1) a progress panel (top left) that shows the status of the underlying AutoAI compute job (which may take from minutes to days to run), 2) a visualization of the AutoAI pipeline (top right) showing the different operations performed on the data and the models produced from each pipeline, and 3) a leaderboard (bottom) that ranks each model that was produced based on a user-specified metric (e.g., ROC AUC). Users are able to click into each model in the leaderboard to see more details.

This demo was used in our interview study to provide informants with a baseline understanding of how an AutoAI tool might be designed: its features, functionality, and interaction design.


\section{Related Work}
We organize related work in the fields of HCI and CSCW into three parts: human intervention in data science, data science teams and disciplinary diversity, and data science tools.

\subsection{Human Intervention in Data Science}
HCI and CSCW researchers have studied how data scientists work with their data (e.g., ~\cite{wang2019,dourish2018datafication,muller2019datascience,neff2017critique,passi2017data,passi2018trust}). Passi and Jackson examined how data science workers deal with rules with doing data science and concluded that most scholars use rules as a basis and framework (``rules-based'') but not as a set of \emph{required} ways-of-working (i.e., not ``rules-bound'')~\cite{passi2017data}. Pine and Liboiron further brought to light the often-invisible measurement plans through which data science workers decide their own rules for what is considered data, how those data are to be represented in a formal repository, and how those data are to be combined~\cite{pine2015politics}. 

More recently, Heer has discussed how portions of the data science process can be automated, while still providing individual agency to data scientists~\cite{heer2019agency}. Heer is a proponent of the use of shared representations -- textual/visual representations of operation specifications, e.g. for data transformation or visualization -- that can be edited by both man and machine. These representations enable both people and algorithms to contribute to a task, as well as learn from the other based on their contributions.

Gil et al. focused specifically on the AutoAI use case and have developed a set of guidelines for the development of AutoAI systems~\cite{gil2019towards}. They used a combination of top-down and bottom-up approaches in analyzing how users currently work with machine learning applications; the former based on their own experiences, and the latter based on analyzing and reconstructing the process of constructing machine learning models as reported in two scientific papers. From their analysis, they developed a set of requirements for human-guided machine learning systems.

Another related research area to the user experience of AutoAI is Interactive Machine Learning. A comprehensive review of this area has been published by Amershi et al. in AI Magazine in 2014~\cite{amershi2014power}. Interactive machine learning research emphasizes building better user experiences for human coders to label data and solicit richer information from humans. Interactive machine learning would serve as a complement to AutoAI, producing labeled data sets for which AutoAI would then be able to build models. In the context of data scientists interacting with AutoAI, we focus more on the process of how data scientists interact with AutoAI systems to create satisfactory models.


Much of the work conducted on the automation of data science has focused on specific kinds of work between human and AI, such as labeling data or using shared representations. Less attention has been placed on the broader implications of what happens to the practice of data science with the introduction of automation. Our work attempts to characterize how data science workers feel about this coming automation and how it will affect the practice of their craft.




\subsection{Data Science Teams and Disciplinary Diversity}
Feinberg described the work of data science as the \emph{design} of data~\cite{feinberg2017design} (see also Seidelin's account of data as a design material~\cite{seidelin2018developing}), and Kiss and Sziranyi noted that datasets in some fields are ``commonly handcrafted''~\cite{kiss2013evaluation}. Patel et al. explored analogous human-constructive activities during feature engineering~\cite{patel2008investigating}. Muller et al. summarized and extended this analysis, concluding that humans may intervene between data and analysis in five ways, according to each data science worker's professional discernment: discovery of data, capture of data, design of data, curation of data, and even creation of data~\cite{muller2019datascience}. In these ways, data science workers engage deeply with data \emph{before} data science modeling activities, \emph{during} model selection~\cite{song2016big,van2014data} and \emph{after} model refinement, such as when testing for bias~\cite{d2017conscientious,walker2015professionalisation}.

Although data science provides powerful tools, there continue to be issues with the required knowledge to apply those tools to specific situations. In a study of urban revitalization, Green et al. concluded in part to ``Apply local knowledge before drawing conclusions''~\cite{green2015mining}. Berrer et al. similarly argued for a need to acquire and incorporate rich domain knowledge in predicting outcomes of football games~\cite{berrar2019incorporating}. From the perspective of domain experts, Bopp et al. described difficulties among non-profit/NGO organizations to make use of data science: small organizations have extensive local and domain knowledge, but they lack the skills to apply data science tools to their problems~\cite{bopp2017disempowered}.

While some scholars hope that more sophisticated abstractions can incorporate domain knowledge~\cite{villanustre2014big}, others are turning to interdisciplinary human teams to combine domain knowledge with data science knowledge. Grappiolo et al. advocate major feature enhancements to include domain experts as effective users of data science tools such as Notebooks~\cite{grappiolo2019semantic}. Passi and Jackson show the complexity of many data science projects and call for the inclusion of diverse expertise in data science teams~\cite{passi2018trust}. Stein et al. describe the needs for complex domain knowledge and multidisciplinary teams in sports data science~\cite{stein2017make}. Borgman et al. ask feminist questions about who provides what kinds of data in data science teams, again arguing for interdisciplinary combinations of data science knowledge with domain knowledge~\cite{borgman2012s}. Viaene said simply, ``It will take a lot of conversation to make data science work. Data scientists can't do it on their own.''~\cite{viaene2013data}.


Data science is a diverse and complex practice, with many stakeholders beyond those that actively work with data or write code. Our work focuses on this latter group -- active practitioners of data science -- in order to build an initial understanding of how they perceive AutoAI changing how they practice their craft. We discuss the need for a more comprehensive view of the impact of AutoAI on data science practice in Section~\ref{sec:limitations}.


\subsection{Data Science Tools}
\label{sec:ds_tools}

The exploratory nature of data science work often uses visual analytics, sometimes in a group setting~\cite{heer2012interactive}. Examples include the early Data Voyager system~\cite{heer2007voyagers}, and the more recent TensorBoard module in TensorFlow~\cite{dang2018predict,martinez2016overview}. These visual analytic tools enable data scientists to more rapidly build an understanding of their data set and develop models for it, but they fall short of providing an automation of these tasks.

The use of Notebooks for organizing data science code (e.g. Jupyter Notebook~\cite{kluyver2016jupyter,web:jupyter}, JupyterLab~\cite{web:jupyterlab,granger2017jupyterlab}, and Google Colaboratory~\cite{web:colab}) has been widely adopted by data scientists. Notebooks provide many of the benefits of Knuth's vision of literate programming~\cite{knuth1984literate} through a stable, structured combination of program code and descriptive text. Many companies, including Netflix~\cite{ufford2018beyond} and IBM~\cite{dobrin_ibm_analytics_2017}, are building their in-house data science infrastructure around the Notebook system, often coupled with other version-control and application deployment systems. 

HCI researchers have explored how users interact with Notebooks and how Notebooks could be modified to improve interactions. Rule et al. showed that people make less use of the textual description features of Notebooks, characterizing the problem as exploration vs. explanation~\cite{rule2018exploration}. Kery et al. looked at this issue in a different way, discovering that data science workers' efforts to construct a narrative in the text features of the Notebook actually led to reduced readability of the resulting Notebook contents, highlighting problems with exploitation of one's own or others' Notebooks~\cite{kery2018story}. The native Jupyter Notebook supports only asynchronous sharing, one Notebook at a time, through repositories such as Github, and data science workers must share their work with low granularity (e.g., not at the micro-level of character as in Google Docs~\cite{dakuo}, but rather at the macro-level of the entire Notebook). The analyses of Kery et al. and of Rule et al. show a series of asynchronous collaboration obstacles around words that begin with ``ex-'': exploration, explanation, and exploitation.


In our study, we examine perceptions of AutoAI technology and question whether it will be perceived as just another tool, like Notebooks, or whether it can take on a greater role in data science practice. This understanding could help shape the future development of AutoAI technologies.

\section{Methodology}
In order to build the understanding of data scientist work practices and their views on AutoAI in the real-world business setting, we conducted semi-structured interviews (following research method in~\cite{dimicco2008motivations,iqbal2010notifications}) with data scientists (N=20) who conduct data science tasks as their primary job function at IBM. The company is so big that the Business Units where these informants are from are very different, thus despite these informants are from the same company, their reported experiences have generalizability to some extend. Nevertheless, we acknowledge the limitation of this sample and we will discuss in the \ref{sec:limitations}.

\subsection{Informants}
As discussed in Muller et al.~\cite{muller2019datascience}, there are a wide variety of roles within data science, ranging from the scientists and engineers who collect and clean data and train and deploy models to leaders and executives who define policies around data governance and ensure use of data conforms to government regulations (e.g. GDPR~\cite{wiki:gdpr}). For our study, we aimed to focus on the former group of people who conduct the technical work of data science. Our criteria for recruitment included: informant has ``data scientist'' in their job title, is actively working on a data science project, and has data science as a primary job responsibility. We used a snowball sampling method, beginning with colleagues who we knew were data scientists, in order to build our sample.


We recruited a sample of 20 data scientists for our study; 11 were female (55\%) and 9 were male (45\%). All of our informants were advanced practitioners of data science who work on concrete business applications of data science. Despite the uniformity of the ``data scientist'' job title, the data science projects they reported and the ways in which they worked on those projects varied. A quarter of our informants worked with other companies on data science projects, collaborating with a wide range of other roles including IT, sales, project managers, business analytics, executives, and other decision makers in industries including telecommunications, banking, insurance, manufacturing, and utilities. Over half of our informants work in healthcare, alongside medical experts and clinical practitioners on projects that often involve electronic medical record (EMR) data. The rest of our informants work on projects such as optimizing the price of a company's products or predicting the failure of hardware or cloud services. Our informant pool ranged in seniority and included junior and senior data scientists, data science managers, and a director of data science. Table~\ref{tab:participants} gives a summary of our informants.

Many of our findings are colored by our informants' unique circumstances, such as working with clients or domain experts who are less technologically sophisticated, working in a regulated industry, or dealing with extraordinarily messy data. We also note that our interviewee pool does not include people in other data science roles, such as citizen data scientists, data scientist novices, executives, or clients of data science teams, all of whom might have different views on how AutoAI may change the practice of data science. We discuss the implications of these limitations in Section~\ref{sec:limitations}.


\begin{table}[ht]
    \centering
    \begin{tabular}{lll}
        Industry / Position     &   Informants \\
        \hline \hline
        Healthcare              &   I2, I5, I6, I7, I8, I9, I10, I11, I12, I15, I16 \\
        Telecommunications      &   I1, I14, I15, I18\\
        Insurance               &   I3, I15, I16 \\
        Retail                  &   I13 \\
        Banking                 &   I17 \\
        Manufacturing           &   I17 \\
        Cloud \& IT             &   I4, I13, I20 \\
        Utilities \& Weather    &   I19 \\
        \hline 
        Senior-level position   &   I1, I3, I7, I13, I14, I15, I20 \\
        Manager / Director      &   I7, I14, I15 \\
        Consultant              &   I1, I11, I13, I16, I17, I18 \\
        \hline
    \end{tabular}
    \vspace{.2cm}
    \caption{Informants work on projects across numerous industries. Some informants work in multiple industries. In addition, some informants have a senior position or are part of management. at various levels of seniority and management.}
    \label{tab:participants}
\end{table}

\subsection{Semi-Structured Interview}
We designed and conducted a semi-structured interview study with simple prompt questions and primarily rely on informants reflection and recall of experiences. The interview script contained three major parts:

\begin{itemize}
    \item We asked the informant to describe their current work practices as a data scientist and reflect on it. Specific topics included informants' job roles, projects, team size, collaborative practices, and business objectives.
    \item We asked the informant's prior familiarity and experience with AutoAI systems and their general attitudes towards them.
    \item We showed an AutoAI system demo (discussed in Section~\ref{sec:autoai_demo} and shown in Figure~\ref{fig:autoai-ui}) to help informants establish a baseline understanding of AutoAI. The demo sparked discussion of the future of data science work. Specific topics include perceived benefits and drawbacks of AutoAI, trust of AutoAI systems and models generated by AutoAI, and how AutoAI may fit into their data science work practices.
\end{itemize} 

Each interview lasted between 35-60 minutes. Due to their semi-structured nature, the topics covered in each interview could vary, and time was spent based on how much each interviewee had to say about each topic. Of the 20 interviews, 4 were conducted in person and the rest were conducted via a videoconferencing system. All interviews were audio recorded and transcripts were produced by a professional transcribing service. In total, we collected about 80 pages of interview notes and roughly 400 pages of transcriptions from 17 hours of interviews. The first two authors of this paper then conducted an open coding process, identifying themes related to data science work practices and perceptions of AutoAI. These themes were discussed and harmonized over several iterations, and specific examples were identified from the source interviews for use in this paper.



\section{Results}
In presenting our results, we first provide context through a discussion of the work practices reported by our informants, including both individual and collaborative practices. We next show the complex, skeptical, grudging, and/or accepting responses of informants to the concept of AutoAI. Finally, we review informants' ideas about their possible future relationships, positive and negative, with AutoAI tools and features.




\subsection{Work Practices of Data Scientists}
All of our data scientists are active practitioners of their craft, conducting work that includes curating and cleaning data sets, writing code, building and evaluating models, and communicating results with colleagues, clients, and other stakeholders. Our informants reported that their biggest challenges lie in the data pre-processing step: finding and curating the data, cleaning the data, engineering features, etc. Data science is practiced as a ``team sport,'' with a high degree of communication and coordination amongst our data scientists and their stakeholders. These findings are consistent with ones reported in previous literature (e.g.,~\cite{muller2019datascience, dobrin_ibm_analytics_2017, heer2019agency, hou2017hacking}), and thus we do not elaborate on them.

\subsubsection{Data Science As A Service}
About a quarter of our data scientist informants work in a business division whose mission is to provide data science consulting services to companies who do not have a well-established data science practice. These consulting data scientists work as a small squad of 2-3 data scientists, and work together with client-side data science teams (if there is one), business owners, sales, IT support, and domain experts. These projects often span either 6 or 12 weeks, and the typical outcome is the creation of a proof of concept (POC) demonstration of how data science (and the associated AI and machine learning techniques) can be used to uncover new insights from the client's data and drive business value. Clients have the option to extend the project into a full product development. One example of a client project was described by I11.

\begin{quote}
    ``One engagement that we finished recently... a classic demand forecasting problem. [The client] incurs charges from the people they're selling their product to if they don't have enough product, and they lose money if they have too much product... They had their own data and they also had external data sources. Can we use this external data and some machine learning techniques to improve our forecasting model?'' (I11)
\end{quote}

One interesting theme that all of our consulting data scientists reported is the educational function of the projects. While developing a data science solution, the data scientists also train members of the client organization.

\begin{quote}
    ``We're educators. That's why we [need to] go in there. [They prefer us to online training] because we're live people where they can talk to us.'' (I1)
\end{quote}

The client companies also want to leverage these data science experts to help them establish a good data science practice of their own.

\begin{quote}
    ``I've been in situations working with clients where they want to adopt data science as a practice and the tools but they don't really know where this would kind of fit into their current business model... It's kind of between like teaching and consulting.'' (I11)
\end{quote}

Some common challenges that informants reported stem from client companies' lack of sophistication in interpreting complex models and results.

\begin{quote}
    ``The best [example to] illustrate this is that when you have a classification problem, there are several models we can choose from, the simplest of which is logistic binary classification. But in general, SVMs or random forests perform way better than logistic regression. But what they all lack is interpretability. With logistic regression, if you print out the coefficients... you can see the importance of those features.'' (I18)
\end{quote}

Lack of data and limited time these informants are given to complete the project can push data scientists to choose models that are easier to interpret over those that are more accurate.
 
\begin{quote}
    ``We get very few clients ...  even ask for a deep learning model ... deep learning models improve as the data gets bigger. And often times, we haven't had clients that are sophisticated enough to have that much data ... the projects they end up doing with us can't be that critical or heavy because of the time restraint... their focus is to see our process, how we do what we do, and to understand what we're doing.'' (I18)
\end{quote}

\subsubsection{Challenges of Industrial Data Scientists}
Many of our informants work on projects in healthcare, insurance, banking, or other highly-specialized industries with sensitive data. These data scientists reported unique challenges with the data produced in their industries.

As reported in many sources, much of the time a data scientist spends is on finding the data and cleaning data to make it ready for modeling.


\begin{quote}
    ``A lot of the work is intensive in acquisition and cleaning... cleaning is very difficult, as we're trying to integrate data from multiple systems, and multiple locations within each system. It gets messy really quickly.'' (I9)
\end{quote}

Many informants expressed difficulties in interpreting healthcare or other domain specific data and asserted the importance of having domain knowledge to be able to clean the data and generate features.

\begin{quote}
    ``... it is very important to work with a Doctor... there are different ways to refer to the same drug and at different quantity, so we are getting some very specific features that really [require] clinical knowledge.'' (I2)
\end{quote}

Also, after building a model, data scientists often need to find a domain expert to help interpret the results and validate whether they make sense or not.

\begin{quote}
    ``You find out those results, then you find the person [domain expert] to ask questions, if it doesn't make sense, you ask more questions [and re-build the model].'' (I2)
\end{quote}

\subsubsection{Collaborative Practices}


Data science is very much a ``team sport''~\cite{dobrin_ibm_analytics_2017}, and the practice of data science in enterprise settings often requires participation from people in a number of roles beyond data scientists and engineers, such as project managers, subject matter experts, clients, and executives~\cite{passi2018trust,muller2019datascience}. While collaborative practices of data scientists have been discussed extensively in the literature (e.g.~\cite{green2015mining,berrar2019incorporating,grappiolo2019semantic,passi2018trust,viaene2013data,stein2017make,borgman2012s}), our informants placed much emphasis on their interactions with subject matter experts, in order to understand the data with which they worked, and on other data scientists, in planning how those data would be cleaned, modeled, and presented to the client. 

Our informants reported collaborating more at the level of ideas rather than at the level of code (i.e. working together with their ``heads'' rather than their ``hands''). They worked with clients to obtain data sets, with subject matter experts to learn about the meanings behind the raw data and define meaningful features, and with each other to strategize about cleaning and modeling the data, and sharing insights discovered from working with the data. The way in which our informants described conducing this work was as a ``scatter-gather'' process, in which they would: 1) have an initial gathering in which the overall goals of the project and the data sets were discussed, 2) conduct separate, individual analyses of the data set, 3) re-gather to discuss insights learned, and 4) repeat steps 2-3 until the project's completion. Input from subject matter experts was sought along the way at the ``gather'' points, but the heads-down work of working with the data and writing code was generally described as being conducted in isolation.

\begin{quote}
    ``... we don't sit and put our hands together on one computer and do it. We might put our heads together to come up with a methodology... and we can do that with the client as well, but coding in and of itself isn't really collaborative.'' (I18)
\end{quote}

The division of labor also happened when trying out different models (e.g., logistic regression vs. SVM) or building different components in the workflow (e.g., feature engineering vs. selecting models).

\begin{quote}
    ``We divide [our] work by models... We have lots of models and we have some of the data processing work commonly shared by other models.'' (I3)
\end{quote}

And from I5,

\begin{quote}
    ``We don't work on a component together. Rather, we work on different components... Rather than three people working on one component, it becomes a bottleneck... so we segregate across components and each one takes one component and tries various methodology for it.'' (I5)
\end{quote}

This way of working impacted how informants thought about AutoAI and where AutoAI might be integrated into their workflow. We explore these implications in the following two sections.

\subsection{Perceptions of AutoAI}
\label{sec:autoai_perception}
Many of our informants had previously heard about AutoAI technologies (11 out of 20), only a few had tried them out once or twice (5 out of 20). Informants reported mixed feedback about those systems' utility and the impact they will have on data science. On one hand, informants felt that AutoAI technologies represented the future of their practice and will help them reduce the amount of time to derive value from data. On the other, informants felt that AutoAI might diminish the technical depth required to perform good data science, leading to situations in which poor-quality models were developed and deployed. Despite both positive and negative perceptions, many informants did feel that the trend toward automation was unavoidable, and they would need to adopt AutoAI technologies into their work practices whether they liked it or not. This sentiment was succinctly captured by I1 and I14.

\begin{quote}
    ``AutoAI is the future. I hate it, but it is the future.'' (I1)
\end{quote}
\begin{quote}
    ``I think it's unavoidable, automation is going to happen.'' (I14)
\end{quote}

In the next several sections, we look more closely at how data scientists thought their work might change.

\subsubsection{Perceived Benefits of AutoAI}
The most valuable aspect of AutoAI is that it can reduce the amount of time data scientists spend going from data to insights. AutoAI provides a starting point from which data scientists can begin working to develop an understanding of their data, including analytic alternatives.



\begin{quote}
    ``Yes, I think [AutoAI] could possibly be useful... especially if AutoAI is fast, it feels like a great starting point... If you could get some defined modeling done using this very fast tool... (I10)
\end{quote}


\begin{quote}
    ``If [AutoAI] will save me time, I can run more of the experiment.'' (I6)
\end{quote}

AutoAI is also perceived to save data scientists time by providing a ``baseline'' measure of how well a model might possibly work on the data, which the data scientist can use to calibrate their own work.

\begin{quote}
    ``AutoML is really just a huge time savings and it gives you a nice baseline that you can then iterate off of.'' (I11)
\end{quote}

\begin{quote}
    ``The first impression is it will do everything for you and maybe you'd get a deeper understanding of the data, a model of what the basic benchmark model looks like.'' (I3)
\end{quote}

As informants reported in the previous section that they often work under time pressure in client engagements, sometimes they may not need the ``best'' model, but rather one that is good enough to meet the client's requirements. This notion was also discussed by Passi and Jackson: ``Perfect solutions were not required -- models simply needed to be `good enough' for the purpose at hand.''~\cite{passi2017data}. In these contexts, AutoAI could be particularly useful.

\begin{quote}
    ``This type of tool can really build just good enough [models] so that we can get more and more models into the workstream, into the business process.'' (I15)
\end{quote}


Other informants hoped that AutoAI could demonstrate good practices in data science. For example, by seeing the code needed to create a random forest model generated by AutoAI, data scientists can learn and improve their own code. This idea is related to the educational function of AutoAI, which we discuss further in Section~\ref{sec:autoai_teacher}.

\begin{quote}
    ``[Generating code] also helps me in the future if I want to do a regression model and I didn't know how to do it. I have example [code] of that [from AutoAI] to practice.'' (I20)
\end{quote}

\subsubsection{Perceived Dangers of AutoAI}
Despite the advantages that AutoAI might provide in bootstrapping the process of building models, many informants expressed a preference for conducting data science work by hand. One reason is that data science jobs often involve educating others to improve their data science skills.

\begin{quote}
    ``We are not a big fan of anything automated. We prefer to do it hands on. So we want to teach the client more machine learning and, you know, looking under the hood rather than using out of the box.'' (I1)
\end{quote}

Despite the reluctance toward adopting automation, the same informant, I1, was resigned to the fact that automated techniques would become part of data science practice, and compared it to the adoption of other new tools and the automation of clothes manufacturing. However, she still expressed skepticism about the quality of what current AutoAI systems could produce.

\begin{quote}
    ``If I have to do things automated, I will have to go with the flow... these days I'm using Slack with emojis, I wasn't using [that] in the past. I'm not against automation. I mean, obviously we're not sewing our own clothes anymore. Not everything you automate is going to be a quality boost [though].'' (I1)
\end{quote}

Another skepticism regarding AutoAI is that its widespread use may weaken or diminish the technical skills of data scientists.

\begin{quote}
    ``I still feel to be a data scientist you need to understand the fundamentals and that includes coding and understanding your data to tell a story. I feel like those essential building blocks could be lost with [using AutoAI]... '' (I7)
\end{quote}

Informants emphasized that they worry more about the future generation of data scientists and how they may incorporate into this future with automated data science tools.

\begin{quote}
    ``I worry that people who weren't trained in [understanding the] mathematical underlying parts. It's easier [for them] to be lazy about applying [automated] models, and the newer people on our team kind of see that... I just worry about depth of the profession...'' (I20)
\end{quote}

Some informants also felt that the primary goal of AutoAI -- producing models with highest prediction accuracy -- did not always match their goals as data scientists in the real world. While sometimes the data scientist's goal is to produce the most accurate model, other times the data scientist aims to understand relationships in the data. In those cases, models are developed as a means to uncover insights in the relationships between those features, rather than as an end in achieving high predictive accuracy.

\begin{quote}
    ``I think you can solve a classification or a regression problem [with AutoAI] ... But another [scenario] is that I want to understand what are the factors that it is showing and why it's associated with my outcomes...  data scientists may not be interested in looking at optimizing predictions, but understanding relationships of features.'' (I7)
\end{quote}

In other cases, the goal of the data scientist is to replicate models already published in literature. These models may not be the best, but they are models that have been validated by domain experts and peer review. In these cases, the features included in these models are based on prior domain knowledge (e.g. clinical knowledge in healthcare), and the value of the model is that it matches existing known theory and/or replicates known practice. Thus, many of the tasks performed by AutoAI around feature engineering and optimization are not applicable.

\begin{quote}
    ``Typically, we're implementing someone else's model. We're standing on the shoulders of literature. This is typically how it is done in healthcare domain... even if your model is better than the paper's model, we would do the paper's model and the hospitals would prefer that.'' (I9)
\end{quote}


One particularly poignant example of how informants felt subject matter expertise would always be important and could not be replicated by AutoAI comes from I12. In healthcare, the domain knowledge of how low-level procedure and billing codes map onto higher-level events such as ``ER visit'' are conveyed from the subject matter expert to the data scientist in order to be able to extract this meaningful event.

\begin{quote}
    ``In claims data, all you'll get is the procedure code [or] the billing code or whatever... and an ER visit might be a whole host of those procedure codes. And so if I have to extract data I need to like write a heck of a lot of SQL code to do that to define an ER visit.'' (I12)
\end{quote}

This sentiment supports the viewpoint of Gil et al., who argue that subject expertise is a critical input in the data science process, even when automated, as automated techniques would not be able to understand the context behind the data~\cite{gil2019towards}.

\subsubsection{Prior Experience with AutoAI}
To develop a concrete understanding of what it is like to actually work with AutoAI, we asked our informants who had told us they had used it before (5 out of 20) to share their experiences. For purposes of this discussion, we will refer to all AutoAI technologies using the generic ``AutoAI'' term, as we do not want to discuss any limitations or benefits of specific AutoAI technologies, products, or services.

I11 applied AutoAI to a problem that he had already solved as a way to benchmark how well AutoAI performed. Ironically, I11 had not used AutoAI since conducting his testing, and questioned why during our interview.

\begin{quote}
    ``I was pretty thrilled with [AutoAI]... there was a problem that was already solved and I wanted to see if it could do better... I just threw my dataset at it and the craziest part was I didn't need to do any real cleansing of the data or any real transformations to the data... And yes in fact it did improve whatever the loss score was over our original model. So then we ended up revisiting that original model and re-coded it based on [the AutoAI's] recommendation. So that was a pretty good experience, which now that I'm saying it out loud, I'm kind of questioning why I haven't like kept using it... I don't know if it just feels like cheating or something, or I feel like the more I use it the more I'm going to put myself out of a job.'' (I11)
\end{quote}

Not all experiences with AutoAI were positive, however. I14 described how she ran into trouble when trying out an immature AutoAI prototype system. Yet, despite this experience, she still liked the idea of automation.

\begin{quote}
    ``I like automation but in this case what I did not like is that sometimes I didn't know why I was getting error messages. For example, I uploaded the data, I started off and I said yes it's a binary classification, et cetera, and it just told me that some errors happened. And then I couldn't go further...'' (I14)
\end{quote}

\subsubsection{Trusting AutoAI}
Trusting the models produced by AutoAI was discussed in many of our interviews. This was an important segment in our interviews because AutoAI technologies are very new and, to our knowledge, no study has yet been conducted to understand how data scientists will come to trust AutoAI models enough to use them in practice.

One major component of trust is transparency~\cite{thelisson2017regulatory, thelisson2017towards}: seeing how AutoAI built a model, what features it used, and what values its parameters were set to. As pointed out by I1, ``one big issue people have is they don't want black box [models],'' as machine learning models are often described~\cite{lipton2016mythos}.

\begin{quote}
    ``I would trust it but I would hope that for whoever does the model that it would provide a tool for [showing] how the model is built and what kind of parameters it generated... I want the pipeline to be automatic but also transparent.'' (I4)
\end{quote}

Seeing inside the model is not the only component of trust; good accuracy and quality of results also matter.

\begin{quote}
    ``What would make me trust it or not is just accuracy of the results.'' (I10)
\end{quote}

Only a few informants felt that it was important to see the underlying code produced by AutoAI to trust it. I5 felt that trust was established by the maturity of AutoAI in its product cycle, with the value of AutoAI stemming from its time savings in providing him guidance in building his own models.

\begin{quote}
    ``If [AutoAI] has gone through testing... I will trust it. I don't want to see the code. Because if I want I can write it from scratch... I would just choose this because it saves me time.'' (I5)
\end{quote}

Establishing trust can be an iterative process between the data scientist and AutoAI. If the AutoAI completely automates the data science pipeline and provides no opportunities for the data scientist to conduct their own analysis or provide feedback into the AutoAI (e.g. feature or model selection), then trust may not be established.

\begin{quote}
    ``Understanding the data dictates doing any type of analysis. If the AI portion of this is providing me with transparency, why they think my data is not clean or why they cleaned it a certain way, before running any models, I think that will be of value. But just saying okay, taking that data, uploading it, running a model, giving me output, I would not trust it.'' (I7)
\end{quote}

Interestingly, this viewpoint mirrors the dilemma that data scientists have often faced when presenting their results to non-data scientists~\cite{jackson2019casting}. Due to a lack of transparency in the process, unfamiliarity with data, or unfamiliarity with data science practice itself, non-data scientists often react with skepticism, confusion, and questions when confronted with a model produced by data scientists. Thus, without a correspondingly high degree of transparency in an AutoAI system, a data scientist may now feel just as ``in the dark'' as a non-data scientist when being so confronted.

Another aspect of trusting AutoAI is getting other stakeholders to trust it as well. For example, while a data scientist may have confidence in a machine learning model, an executive may be less keen on actually deploying it in their business. Although many technology companies have built very successful businesses on top of AI and machine learning, there still seems to be reluctance in some industries to adopt AI.

\begin{quote}
    ``[AutoAI] is very interesting to me... The only thing I would see as a problem is translating this to whoever the business person is. I've seen  VPs at other companies who... don't trust machine learning as it is or AI." (I12)
\end{quote}

To equip data scientists with means to make business cases for the adoption of AI (broadly) and AutoAI (in specific), informants desired AutoAI to provide features to support the telling of stories from their data. Informants suggested one way to do this is to focus on specific data points and craft stories for how the model generated predictions for those data.

\begin{quote}
    ``What I found really powerful when presenting models to people is showing them the actual, real data points. So-and-so is age 25 who runs at North Lake Park every month and hasn't missed her run for the last five years... That's why she buys more shoes. As soon as you show that, then people are like, `Oh, yes, totally. We trust you [and your model] now.' '' (I13)
\end{quote}


\subsection{AutoAI Collaborate with Data Scientists}
\label{sec:autoai}
Given that informants expressed a mixed range of feelings toward the utility of, quality of, and desire to use AutoAI technologies, coupled with strong feelings of inevitability, we sought to understand how they felt AutoAI would manifest in their work practices: do they feel it will integrate with human effort into the different stages of the data science pipeline, or do they believe it will entirely replace a data scientist's job?

Our informants believed the role of AutoAI will be dichotomous, augmenting human effort in some stages of the data science pipeline and reducing it in others. Further, AutoAI was perceived to be capable of providing tangible benefits beyond its intended purpose of building highly accurate models quickly, in that it can also \emph{educate} data scientists (and aspiring data scientists) about how to practice their craft. The data science job will still exist, but the role of a data scientist may shift as non-data scientists become empowered through AutoAI. This notion was best articulated by I7.


\begin{quote}
    ``[AutoAI] will definitely make an impact... I think the domain expert will still be there, but I think over time the role of the data scientist could get a bit murky... [would there] actually [be] a need for a [separate] data scientist when a subject matter expert could do [data science] themselves?'' (I7)
\end{quote}

\subsubsection{AutoAI as Collaborator}
\label{sec:autoai_collaborator}

Many informants discussed how AutoAI can provide a useful starting point when working with a data set, from which they can conduct further exploration. This view of AutoAI as a \emph{collaborator}, akin to their data scientist colleagues today, entails automation of the high-effort, tedious, or unattractive portions of the data science pipeline.

\begin{quote}
    ``I think data scientists would start using the [AutoAI] tool...  and then they might tweak it and add things.'' (I15)
\end{quote}

Human data scientists could divide labor by phases of the data science workflow, such as having AutoAI focus on data pre-processing while humans perform modeling.

\begin{quote}
    ``AutoAI or AutoML would be most useful in the data prep phase.'' (I12)
\end{quote}

AutoAI can also act as a first-order collaborator within a data science team. For example, our informants described a ``scatter-gather'' approach to their collaborative work. During the ``scatter'' phase, AutoAI
can improve the productivity of an individual data scientist by automating aspects of data cleaning, feature engineering, and model building -- tasks that today's AutoAI systems currently address. For example, AutoAI can provide a baseline for which kinds of models might produce the best results. 

\begin{quote}
    ``It's like if I have a training set and I want to find out which is the best model for a classifier, I can give it and [AutoAI] will give me the best model and the best settings and I am very happy with that.'' (I5) 
\end{quote}

\begin{quote}
    `` [AutoAI] gives me a sense of what types of models might be performing better than others... I can narrow down to the ones I might want to focus on or have my team dig up and work on each model individually... I can...  focus more on reporting the results.'' (I20)
\end{quote}

However, there are also opportunities for infusing AutoAI into team collaboration practices. AutoAI can aid in the creation of an analysis plan by providing specific recommendations -- analyses to be performed, data to be labeled, features to be examined, models to be evaluated -- to the human members of a data science team based on that team's needs and expertise. For example, our informants strongly felt that AutoAI would never replace the need for subject matter expertise. Imagine, then, an AutoAI capable of identifying and requesting assistance from such experts when confronted with a need for semantic labels or relating data items across multiple tables?

AutoAI could also steer itself toward performing analyses that other team members haven't yet tried, by monitoring their efforts and identifying gaps during the ``gather'' phases of work. It can also act as a \emph{competitor} with human data scientists, offering challenges to improve upon the baseline models it generates. Mixed-initiative group communication settings, such as team chat spaces, may be an apt setting for enabling this kind of human-AI collaboration.

AutoAI may also be a valuable collaboration partner for people who are not trained in data science (e.g. citizen data scientists, domain experts, or line of business users), by providing no-code solutions for quickly building models.

\begin{quote}
    ``I think [AutoAI] might be easier for [domain experts], they might not be able to code'' (I7)
\end{quote}

\begin{quote}
    ``If you don't know any data science then [AutoAI] has a lower bar, everyone can use it which is very good. However, you lose your flexibility.'' (I3)
\end{quote}

However, some skepticism was expressed in how to interpret models and report the results generated by AutoAI, if one has not had hands-on experience with a model.

\begin{quote}
    ``How can a data scientist be able to understand the output of these results if they cannot code and have never been hands-on before?'' (I7)
\end{quote}

In contrast, a collaborative AutoAI may enable non-data scientists users to collaborate with human data scientists and make meaningful contributions. In this way, AutoAI is perceived as a scaffold that holds up the contributions of the non-data scientist when fundamental knowledge of data science is lacking.

\begin{quote}
    ``To the extent that I am collaborating with others who might not be hard core coders, the ability for them to fire up [AutoAI] lets them make fairly meaningful contributions...'' (I11)
\end{quote}

One requested feature from informants for a collaborative AutoAI is the ability to customize features that should be included in a model. This ability might be especially important for including the expertise and feedback from subject matter experts into the system, especially in situations where specific domain knowledge is critical for building a sound model.

\begin{quote}
    ``It seems like you put in a data set then it will do it for you, but if I want certain features to always be in my model, is there a way to provide that flexibility?'' (I7)
\end{quote}

However, informants expressed skepticism of whether the entire data cleaning and feature engineering processes could be performed by AutoAI, especially when deep subject matter expertise was required. These informants felt that a human must remain in the loop to constantly provide input during these stages.

\begin{quote}
    ``One of the most predictive variables that came out of [the Titanic data set] is if you... extracted the surname as a feature, that was extremely predictive in whether somebody survived or not... [T]here has to be some intuition to how you're treating that categorical variable, and then it's up to the data scientists, their creativity, to add more value to the model based on their business logic.'' (I13)
\end{quote}

Another aspect of AutoAI acting as a collaborator is in the production of interpretable models. Often, a data scientist's job is to explain to other, non-data scientist stakeholders, the meaning and the story behind their models. The ability to interpret models is crucial, and models that are overly complex may not be as readily accepted. Thus, there may be a tradeoff between predictive performance and interpretability, which the data scientist must navigate depending on the needs of their stakeholders. Informants desired a feature in AutoAI to incorporate interpretability as a business requirement for the models it produced.

\begin{quote}
    ``In a lot of these clinical applications, there's interest in interpretability. So, if one model is much simpler to run, or if some linear combination that would be trivial for a doctor to do perhaps even in their head, I'm going to give priority to that model.'' (I10)
\end{quote}

\subsubsection{AutoAI as a Teacher}
\label{sec:autoai_teacher}

Another role for AutoAI is being a \textit{teacher}. Multiple informants reported that data scientists often need to educate others about data science, and it would be helpful if AutoAI could help fulfill that task. However, numerous concerns were raised in our interviews about AutoAI being used by people who are not trained data scientists, making it ``too easy'' to build models that might be of questionable quality; for example, a non-data scientist might not know how to interpret different accuracy metrics, or might not know about the dangers of overfitting a model to the training data. 

\begin{quote}
    ``I guess my only drawback of this [AutoAI] tool is I could see it being used by like someone who doesn't really know what they're doing. You know, put it in the hands of just a business analyst who doesn't really know much about machine learning and just quickly produce something, thinking that's it.'' (I12)
\end{quote}

Viewing AutoAI as a teaching tool may alleviate these concerns. For people without a strong data science background, such as citizen data scientists or line of business users, AutoAI can explain its reasoning for why it built models or engineered features the way it did. Answers to these ``why'' questions about the data science pipeline are crucial for being able to establish trust in a model, and for these users to learn. Many tools and techniques are being developed in the area of Explainable AI (XAI), focused on areas such as feature importance~\cite{friedman2001greedy} and interpretable predictions~\cite{Ribeiro_2016}. I6 discussed the importance of building explainable models and how it can sometimes be difficult, especially when clients don't understand the underlying modeling techniques.

\begin{quote}
    ``But then [with] deep learning... unless we can clearly explain what it means, clients will never buy it... Explainability is always a bottleneck for us. So whenever you can define a model, it is important.'' (I6)
\end{quote}

XAI techniques can be integrated with AutoAI user interfaces to provide answers to some of the ``why'' questions, although new techniques may need to be invented to provide other kinds of explanations (e.g. what the engineered features are and how particular ones were chosen). Another idea was raised by I13, who suggested that AutoAI cite its sources to justify the decisions it made.

\begin{quote}
    ``If I were using AutoAI... to select a model architecture for me... if there's education or some feedback loop like, `hey, your data's distributed like this, the research shows that really wide layers perform better than really deep architectures'... If it [can provide] a citation of a paper and why we performed some task for you automatically. I would love that.'' (I13)
\end{quote}

Another aspect of AutoAI as a teacher is that it can be used to demonstrate best practices. By supplying the code that generated a model, AutoAI can both demonstrate how it achieved its results and provide in-context explanations for its modeling choices (e.g. by describing alternatives it tried). The generated code would also serve as an artifact that captures the \emph{how} of data science, which may be useful both for non-data scientists to learn the field, and also for data scientists who may have forgotten specific details of a task, such as training a rarely-used model.

\begin{quote}
    ``[Generating code] also helps me in the future if I want to do a regression model and I didn't know how to do it. I have examples of that [from AutoAI] to practice.'' (I20)
\end{quote}


\subsubsection{AutoAI as Data Scientist}
The obvious role of AutoAI is to completely automate every stage of the data science pipeline. Data scientists are expensive and not every organization possesses the budget or technical capabilities to develop their own data science practice. Thus, there may be tremendous business value in being able to replicate as much as possible of the data science profession with AutoAI.

\begin{quote}
    ``Who wants to keep paying data scientists \$200,000 a year to go do the same steps over and over again and not be high maintenance? Of course, the data science roles are going to change... I do think the tools will automate a lot. They'll be less data scientists needed than today, and they'll be focused on higher value work. So, I do think the data scientists should be part of a Center of Excellence or R\&D team, and then the folks inside the line of business need to be... like citizen data analysts.'' (I15)
\end{quote}

One downside to AutoAI acting as a data scientist is that much of the ``pain'' that data scientists endure in cleaning data, engineering features, and building models might actually be beneficial for developing deeper insights into the data. By actively exploring, probing, and modeling a data set, the data scientist builds their own mental models of what features are important and how they are related. Entirely automating this process might eliminate a data scientist's ability to deeply understand their data set; thus, there may be value in the model-building process that is lost with AutoAI.

\begin{quote}
    ``There's a lot learned in the model building process that you wouldn't learn if you just threw everything at [AutoAI]... [T]he information [learned] along the way is super valuable. How would you capture that if all that's automated?'' (I13)
\end{quote}

In general, informants felt that AutoAI would \emph{enhance}, rather than \emph{eliminate} their jobs. This notion was best expressed by I20 and I14, who were asked about their worries about being replaced by AutoAI.

\begin{quote}
    ``No. [AutoAI] would make my job easier. What would replace my job is if that investigative part, figuring out what even needs to go into auto [AI] tools, [is automated].'' (I20)
\end{quote}

\begin{quote}
    ``As for job security I'm not that worried right now... you will still need people because people will understand the problem, people will understand what solution needs to be put in place... data scientists will be able to focus on something else, but probably are not focused so much on [model building].'' (I14)
\end{quote}

\section{Discussion}
As many AutoAI design features discussed by our informants have already been presented, we focus our discussion on the two ways we identified that AutoAI may manifest itself:  augmenting a data scientist to enhance their productivity, or acting as a data scientist in its own right.

\subsection{The Augmented vs. Automated Data Scientist}
One theme that came up in our interviews, especially amongst informants working in healthcare, was the importance of subject matter experts. The general belief amongst our informants is that there will always need to be some level of human input feedback given to an AutoAI system because it simply cannot infer the relevant subject matter expertise pertinent to understanding the meaning of data in a data set. One concrete example of this idea comes from I12's comment about how low-level procedure and billing codes map to higher-level events such as ``ER visit.'' Our informants expressed much skepticism about AutoAI ever being able to automatically infer this kind of domain knowledge without human input. This finding echoes the conclusions of Green et al. in which the application of local knowledge ought to be applied before drawing conclusions~\cite{green2015mining}.

In this viewpoint, AutoAI serves more to augment the skills and abilities of the data scientist, or of other practitioners attempting to apply data science to their field. The difficulties encountered by NGO workers in applying data science tools to their problems reported by Bopp et al.~\cite{bopp2017disempowered}, for example, might have been alleviated with the availability of AutoAI tools. Thus, the role of the data scientist may shift more toward eliciting relevant domain knowledge from subject matter experts in order to provide guidance to the AutoAI for how to conduct data cleaning and transformation operations, and away from other tasks such as producing and evaluating predictive models, which might be more easily automated. But, there is clearly still a human data scientist in the loop, guiding the actions of AutoAI to produce outputs that satisfy business goals. The roles identified through our interviews -- AutoAI as collaborator and teacher -- fit nicely into this human-in-the-loop model.

One potential downside of the augmented data scientist viewpoint is the concern articulated by I13, that by automating portions of the data science pipeline, data scientists may either lose touch with their craft (e.g. their skills atrophy from lack of use), or they may miss seeing potentially valuable insights in their data (e.g. because they no longer need to actively explore and probe their data set in order to produce quality models). 

We believe that these concerns can be alleviated in part by making explainability a core design goal of AutoAI interfaces. If AutoAI designers take the perspective that the goal of AutoAI is to \emph{teach}, then the operations performed by AutoAI should not be hidden; rather, they should be readily available for inspection, inquiry, and modification.

As a concrete example, if AutoAI determines that a certain model best fits the data scientist's requirements, it should be able to explain why (e.g., ``I chose a random forest regressor because it had the highest accuracy'' or ``I chose logistic regression because it has high explainability and performed within accuracy requirements''). It should also be able to describe which models were or were not considered and why (e.g. ``Random forests were not considered because they can't perform extrapolation''), and it should give the data scientist full control over the final choices for all aspects of the pipeline (e.g. by allowing them to modify the code that implements the pipeline). In addition, if AutoAI can explain its own actions (as discussed in Sections~\ref{sec:autoai_collaborator}-\ref{sec:autoai_teacher}), then we may need to revisit the data science tools discussed in Section~\ref{sec:ds_tools}, adding features to support both AutoAI explanation and human interpretation in a discussion of choices and selections among alternative AutoAI outcomes. Because of the need for interdisciplinary knowledge in data science teams~\cite{grappiolo2019semantic,passi2018trust,stein2017make,viaene2013data}, it will be necessary to design these interpretation and discussion features for users from diverse backgrounds.

In contrast to the augmented data scientist, the viewpoint of ``AutoAI as data scientist'' is one in which AutoAI completely automates the job of a data scientist. This viewpoint was perhaps more attractive to managers (e.g. I15) because of the cost savings associated with needing to hire fewer workers. However, many informants were skeptical that AutoAI would ever be able to completely replicate a data scientist. 

Because of our informants' confidence that AutoAI technologies would act in collaborative capacities with their users, especially with regard to the need for human subject matter expertise, we feel that AutoAI systems should be developed with a mindset of \emph{augmenting} rather than \emph{automating} human capabilities across the heterogeneous needs of data science teams. Heer argues for the need to ``integrate proactive computational support into interactive systems,'' and gives many examples of how AI technologies, combined with good UX design, can be used to create systems that automate tasks while preserving human agency~\cite{heer2019agency}. Gil et al. echo the sentiments expressed by our informants on the importance of including subject matter experts in the data science process, saying ``without this knowledge, machine learning models might optimize model search criteria but be inconsistent with what is known about the data and its context.''~\cite{gil2019towards}. Li gives an example of how AI automation should focus on enhancing the strengths of humans ``like dexterity and adaptability'' by ``keeping tabs on more mundane tasks and protecting against human error, fatigue and distraction''~\cite{li2018goodai}. Our view extends the recent popularity of these ``human-centered'' approaches to AI toward more collaborative possibilities of mixed initiative between human workers and AI agents.

As discussed earlier, if human data scientists know when to be rule-\textit{bound} and when to be rule-\textit{based}~\cite{passi2017data}, then how does that discernment translate to AutoAI? Similarly, if data science workers often design, curate, or create their data~\cite{muller2019datascience}, should we consider these interventions as necessary acts of human judgment, or should we instead rely on AutoAI to remove those human actions? We also note that the activities of creating and enforcing a measurement plan~\cite{pine2015politics} are often invisible. At this time, it is not clear how AutoAI can create, support, or even understand the issues in writing and updating a measurement plan, and it is not clear how AutoAI would be able to decide when to apply that measurement plan in a rule-bound way or when to make intelligent deviations from that plan in a rule-based way~\cite{passi2017data}. We may find ourselves returning to Borgman's et al.'s necessary question about data science collaborations, ``Who's got the data?''~\cite{borgman2012s}.


\subsection{Limitations and Future Directions}
\label{sec:limitations}

Our informants were all recruited from our company -- a large, multinational technology company -- and their views may not be fully representative of the larger population of practicing data scientists, or of the broader class of data science workers. One example of how our results might be skewed comes from the fact that almost all of our informants worked in small teams, typically of 2-3 data scientists, with possible additional members in other roles such as engineers, subject matter experts, or client stakeholders. Many interviewees described their collaborative practices as operating on a conceptual level (e.g. discussing approaches to cleaning or modeling data) rather than on a concrete artifact level (e.g. sharing code). Thus, the metaphor we identified as to how their work \emph{happens} is ``scatter-gather.'' The extent to which this metaphor generalizes beyond our informants, and beyond our own organization, is both an open question and an important one for understanding how AutoAI might be integrated into data science workflows. For example, current AutoAI systems focus on empowering the individual data scientist, helping them navigate amongst the multitude of options for processing data and building models. However, what features would an AutoAI system need to fit into a more collaborative environment where code, data, and models are all shared amongst larger numbers of people? Would AutoAI be just another team member, making active contributions to the team's growing base of code? Or should its role be more like a coach or advisor, providing advice and suggestions for the team based on the team's collective efforts (e.g. ``I see Brian tried an SVM, but if you use a random forest you can increase accuracy by 4\%'')? From our study, we begin to see that AutoAI can take on different, distinct roles -- teacher, collaborator, data scientist -- and we see this rather a new promising research direction than limitation, and we look forward to understanding how else it might integrate into data science practice.

Another consequence of our sample of data scientists is that the opinions of other kinds of data workers (e.g. machine learning engineer) and stakeholders (e.g. executives, project managers, clients) is not represented. How these people perceive AutoAI is another open question for additional study. For example, would a client accept a model produced by AutoAI when they feel they are paying for human effort and expertise? How does a project manager account for the lengthy amounts of time an ``AutoAI collaborator'' might take to produce results? Will executives make cuts to data science teams given the automation capabilities of AutoAI, and if so, what new work will existing data scientists perform? But it requires extensive efforts to answer these questions, which is beyond this paper's focus on "data scientists", which we plan to explore in the future.

AutoAI is a new technology, and one that most of our informants (75\%) had not previously used. Because of this novelty and unfamiliarity, many of our informants were only able to provide their perceptions of how it might affect their practice of data science, rather than concrete examples of how it affected their practice, and we believe this is the case for data scientists in general. As AutoAI technologies mature and become more pervasively adopted, additional research is needed to judge exactly how it has affected data science practice, both at the individual and group levels. One aspect that is crucial to adoption of AutoAI is \emph{trust}: how can data science workers and stakeholders come to trust the results produced by AutoAI? We believe that designs that cast AutoAI as ``teacher'' and ``collaborator'' will be helpful in establishing trust by demonstrating knowledge and helpfulness, and we are eager to see how such designs manifest as AutoAI technologies evolve.

\section{Conclusion}
We have presented results from a semi-structured interview study of 20 data scientists who work at a large, international technology company. They reported mixed perceptions of automated AI (AutoAI) technology. While seeing benefits in speeding up the process of building data science pipelines, they expressed concerns around how AutoAI might lower the bar too much with regard to the technical skill required to conduct good data science.

We believe that AutoAI systems ought to be crafted with a mindset of \emph{augmenting}, rather than \emph{automating}, the work and work practices of heterogeneous teams that work in data science. We have uncovered several potential indirect benefits of AutoAI, beyond simply eliminating select aspects of the work of conducting data science. For example, AutoAI can be used to teach aspiring data scientists the best practices of their craft. It can also act as a first-order collaborator in data science teams by making recommendations for analyses and offering alternatives that human team members can discuss and improve upon, giving judicious explanations behind its reasoning to build trust. In these ways, we envision AutoAI as becoming an invaluable new partner for data science teams.



\bibliographystyle{ACM-Reference-Format}
\bibliography{bibliography}

\end{document}